\documentclass[prb,showpacs,twocolumn,superscriptaddress,amsmath,amssymb,floatfix]{revtex4-1}
\usepackage{graphicx}
\usepackage{bm}
\usepackage{bbm}
\usepackage{color}
\usepackage{color}
\usepackage{physics}
\usepackage{bbold}
\usepackage{comment}
\usepackage{footnote}
\usepackage{epstopdf}
\usepackage{color, colortbl}
\usepackage[table]{xcolor}
\usepackage{wrapfig}
\usepackage{hyperref}
\usepackage{upgreek}
\usepackage{appendix}
\usepackage[applemac]{inputenc}

\begin{document}
\title{Programmable two-qubit gates in capacitively coupled flopping-mode spin qubits}
\author{Jorge Cayao}
\affiliation{Department of Physics and Astronomy, Uppsala University, Box 516, S-751 20 Uppsala, Sweden}
\affiliation{Department of Physics, University of Konstanz, D-78457 Konstanz, Germany}
 \author{M\'{o}nica Benito} 
  \author{Guido Burkard} 
\affiliation{Department of Physics, University of Konstanz, D-78457 Konstanz, Germany}

\date{\today} 
\begin{abstract}
Recent achievements in the field of gate defined semiconductor quantum dots reinforce the concept of a spin-based quantum computer consisting of nodes of locally connected qubits which communicate with each other via superconducting circuit resonator photons. In this work we theoretically demonstrate a versatile set of quantum gates between adjacent spin qubits defined in semiconductor quantum dots situated  within the same node  of such a spin-based quantum computer.
The electric dipole acquired by the spin of an electron that moves across a double quantum dot potential in a magnetic field gradient has enabled strong coupling to resonator photons and low-power spin control. Here we show that  this 
 flopping-mode spin qubit also provides with the tunability to program multiple two-qubit gates. Since the capacitive coupling between these qubits brings about additional dephasing, we calculate the estimated  infidelity of different two-qubit gates in the most immediate possible experimental realizations.
\end{abstract}
\maketitle
\section{Introduction}
\label{section0}
The electron spin  is a promising candidate for a qubit \cite{loss2,PhysRevB.59.2070,doi:10.1146/annurev-conmatphys-030212-184248,Awschalom2013} with  well controllable initialization  and   long coherence times,~\cite{Petta2180,RevModPhys.85.961} e.g. of the order of seconds in Si,\cite{Tyryshkin2012} which represent part of the necessary requirements  for building a quantum computer \cite{DiVincenzo1997}. Furthermore, for the realization of any quantum algorithm it is also indispensable to perform multiple operations, unitary transformations known as quantum gates, on one and two  qubits from  a universal set of quantum gates.\cite{PhysRevA.51.1015}
For instance, the controlled-not gate (CNOT) \cite{PhysRevA.52.3457} and $\sqrt{\text{SWAP}}$ \cite{loss2} are examples of two-qubit gates that, together with single-qubit rotations, enable arbitrary multiqubit operations.

The two-qubit gates are often the most challenging since they require coupling between qubits.  A wide variety of coupling types, from direct electron-electron interactions \cite{PhysRevB.75.085324,Taylor2005,Li2015,Ward2016,Shinkai2009,Shulman202,PhysRevB.92.235301,PhysRevB.59.2070,PhysRevLett.111.050503,Veldhorst2015,Nichol2017,Watson2018,Zajac2017} to interactions mediated by the substrate or other intermediate system, \cite{PhysRevX.2.011006,PhysRevX.3.041023,PhysRevLett.111.060501,Delbecq2013,Nicoli2018,Woerkom2018,Borjans2019} have been demonstrated  for qubits defined in semiconductor quantum dots (QDs).   
The first type of  coupling schemes, including capacitive coupling and exchange, are usually short range and would constitute the fundamental ingredient for qubit operations within a quantum computer node. Such nodes can then be joined via interconnects such as superconducting cavities to relax geometric constraints within the architecture.~\cite{Taylor2005,Kimble2008} 

\begin{figure}[!t]
	\centering
	\includegraphics[width=.45\textwidth]{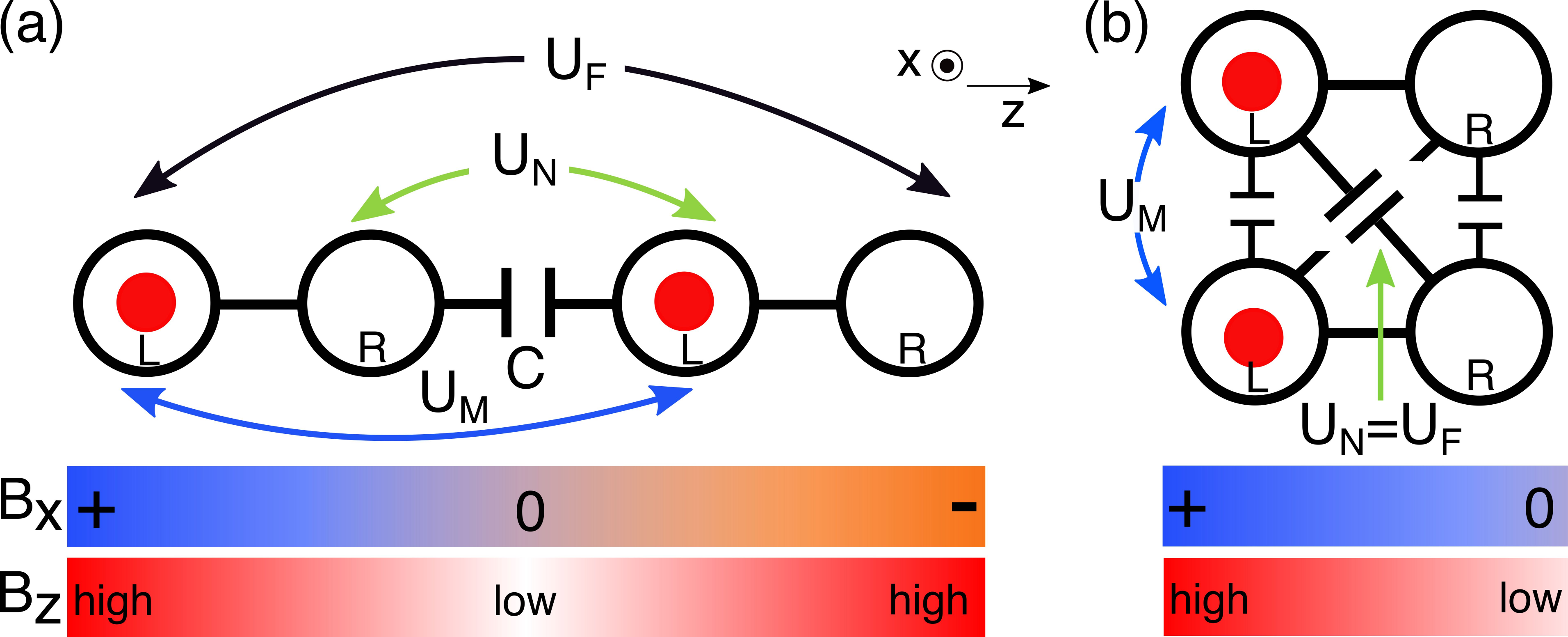}
	\caption{Schematics of two flopping-mode spin qubits coupled capacitively via the Coulomb interaction (C). Each flopping-mode spin qubit consists of a single electron (red dot) occupying a double quantum dot (DQD). The Coulomb interaction between electrons in different DQDs depends on the electron location: The configuration with both electrons  in the inner (outer) dots is associated with a ``near'' (``far")  repulsion energy  denoted as $U_N$  ($U_{\rm F}$), while the value $U_{\rm M}$ describes the configuration with both electrons located either in the left (L) or the right (R) QD.  (a) Horizontal geometry. (b) Vertical geometry. The color bars in both cases represent the intensity of the in-plane ($B_z$) and out-of-plane ($B_x$) inhomogeneous magnetic fields.}
	\label{Fig1}
\end{figure}

Motivated by recent experimental work on  capacitive coupling between QD qubits \cite{Shulman202,Li2015,Nichol2017,Neyens2019,Gao2020} and in quantum dot arrays with different spatial configurations,~\cite{Mortemousque2018,Uditendu2018,Volk2019,Sigillito2019,Kandel2019} we consider here the capacitive coupling between flopping-mode  spin qubits~\cite{Hu2012,Srinivasa2013,Benito2019b,Croot2019}
and investigate the realization of two qubit gates.
As shown in  Fig.\,\ref{Fig1}, each flopping-mode spin qubit consists of a single electron trapped in a double quantum dot (DQD) embedded in an inhomogeneous magnetic field as e.g. provided by a micromagnet, such that the spin qubits acquire an electric dipole that allows them to interact with each other via Coulomb interaction. 
We consider two geometries which can serve as guiding units for  QD networks and scalable systems for quantum computation.
By calculating the Makhlin invariants,\cite{Makhlin2002,PhysRevA.67.042313,PhysRevB.91.035301,PhysRevB.92.125409} from the physical Hamiltonian for the qubit-qubit interaction, we find that multiple two-qubit quantum gates can be realized
in a single step with fidelities higher than 95\%, including the effect of  dephasing.
The micromagnet induced electric dipole has been proven useful to
 integrate this qubit in
 circuit quantum electrodynamics architectures, \cite{Benito2017,Mi2018,Samkharadze2018,Borjans2019} which  allows to interconnect the nodes via electromagnetic fields. 
 Our findings therefore demonstrate the versatility of this type of quantum dot node.

The remainder of this paper is organized as follows.
In Sec.\,\ref{section1} we derive an effective  qubit-qubit Hamiltonian within a perturbative approach in the  Zeeman field gradients, where the couplings exhibit high control via the system parameters, which allows a switch on and off at will. In Sec.\,\ref{section2},  we analytically calculate the Makhlin invariants\cite{Makhlin2002,PhysRevA.67.042313} and demonstrate that the effective qubit-qubit Hamiltonian allows the realization of multiple two-qubit gates in a single step, including CNOT, which, together with single-qubit rotations, has proven to allow universal quantum computation.  Furthermore, by calculating the two-qubit gate fidelity in Sec.~\ref{Fidelity}  we  analyze the robustness of the obtained  gates accounting for dephasing.  After briefly analysing the results for an alternative, simpler geometry in  Sec.~\ref{alt}, we present our conclusions in Sec.\,\ref{conclusions}. For completeness, we provide all the details on the derivation of the analytical calculations reported in this work in Appendices~\ref{sec:diagonalization} and~\ref{sec:SW}.

\section{Model and low energy Hamiltonian}
\label{section1}
We consider two different geometries of capacitively coupled DQDs, one horizontal and one vertical, where each of the two DQDs ($i=1,2$) hosts one electron in its left (right) QD, described as the quantum state $|L^{i}\rangle$ ($|R^{i}\rangle$), as schematically shown in Fig.~\ref{Fig1}. Each of the two geometries can serve as a building block for networks of capacitively coupled DQDs.
In both cases, the Hamiltonian for the capacitively coupled DQDs assumes the form
\begin{equation}
\label{eq:H1}
H=\frac{1}{2}\sum_{i=1,2}\Big[\varepsilon_{i}\tau_{z}^{(i)}+2t_{c}\tau_{x}^{(i)}\Big]+H_{\rm Z}+H_{\rm C} , 
\end{equation}
where the first term in square brackets corresponds to the Hamiltonian for each DQD, the second to the Zeeman field contribution, and the third term to the Coulomb interaction that represents the capacitive coupling between the two DQDs. Here,   $t_{c}$ characterizes the interdot tunnel coupling, $\tau_x^{(i)}=|R^{i}\rangle\langle L^{i}|+|L^{i}\rangle\langle R^{i}|$, $\varepsilon_{i}$ is the energy detuning,  and  $\tau_z^{(i)}=|L^{i}\rangle\langle L^{i}|-|R^{i}\rangle\langle R^{i}|$.
The Zeeman contributions $H_{\rm Z}$  are given by
\begin{equation}
\label{Zeeman1}
\begin{split}  
H_{\rm Z}&=\frac{1}{2}\sum_{i=1,2}\Big[\tilde{B}_{z}\sigma_{z}^{(i)}+(\mp 1)^{i-1}\tilde{b}_{z}\sigma_{z}^{(i)}\tau_{z}^{(i)}\\
&+(\mp 1)^{i-1}\tilde{B}_{x}\sigma_{x}^{(i)}
+
\tilde{b}_{x}\sigma_{x}^{(i)}\tau_{z}^{(i)}\Big]\,,
\end{split} 
\end{equation}
where the $\mp$ sign arising in the second and third terms correspond to the horizontal and vertical configurations represented in Figs.\,\ref{Fig1}(a) and \ref{Fig1}(b), respectively, and $\sigma^{(i)}$ are the spin Pauli matrices. 
The homogeneous and inhomogeneous parts of the $x$ ($z$) components of the external magnetic field are
denoted by $\tilde B_x$ ($\tilde B_z$) and $\tilde b_x$ ($\tilde b_z$). Rotating the spin quantization axis in each dot, this can be expressed via a Zeeman field $B_{z}=\sqrt{\tilde{B}_{z}^{2}+\tilde{B}_{x}^{2}}$ equal in both QDs, a longitudinal  magnetic field gradient $b_{z}=(\tilde{b}_{x}\tilde{B}_{x}+\tilde{b}_{z}\tilde{B}_{z})/B_{z}$, and  a transverse one, $b_{x}=(\tilde{b}_{x}\tilde{B}_{z}-\tilde{b}_{z}\tilde{B}_{x})/B_{z}$ (see Appendix~\ref{sec:diagonalization}).

The Coulomb interaction between electrons in different DQDs, written in the same two-qubit basis as Eq.\,\eqref{Zeeman1}, reads
(see e.g. Ref.\,[\onlinecite{Neyens2019}])
\begin{equation}
H_{\rm C}=-\frac{U_{\rm N}-U_{\rm F}}{4}\tau_z^{(1)}+ \frac{U_{\rm N}-U_{\rm F}}{4}\tau_z^{(2)}+
g\tau_{z}^{(1)}\tau_{z}^{(2)}\,,
\end{equation}
where $g=(2U_{\rm M}-U_{\rm F}-U_{\rm N})/4$ represents the coupling strength and is determined by the value of Coulomb interaction terms $U_{\rm N, M, F}$, which correspond to the Coulomb repulsion between electrons that are near (N) (electron in DQD 1  is in right QD and electron in DQD 2  is in left QD), at a medium (M) distance  (both electrons are either in the right or in the left QD), or far apart (F) (electron in DQD 1  is in left QD and electron in DQD 2  is in right QD) from each other as indicated in Fig.~\ref{Fig1}. 

For simplicity, in the following we consider only the detuning value $\varepsilon_2=-\varepsilon_1=-(U_{\rm N}-U_{\rm F})/2$, which is the most symmetric point.
Note that in the geometry represented in Fig.\,\ref{Fig1}(b) $U_{\rm N}=U_{\rm F}$.
Explicit values of $U_{\rm N, M, F}$ can be obtained by performing two-particle Coulomb interaction integrals as in Refs.~\onlinecite{PhysRevB.75.085324,PhysRevB.91.035301};  in experiments they are directly accessible but depend on the geometry and gates  and are necessarily renormalized due to unavoidable screening effects.\cite{Neyens2019} Hence, below we consider values of the inter-DQD coupling $|g|\sim15-30\,\mu\text{eV}$ as reported in recent experiments.~\cite{Neyens2019} Notice that we assume that any exchange coupling is strongly suppressed due to the absence of tunneling between the two DQDs. 

\begin{figure*}[!ht]
	\centering
	\includegraphics[width=.98\textwidth]{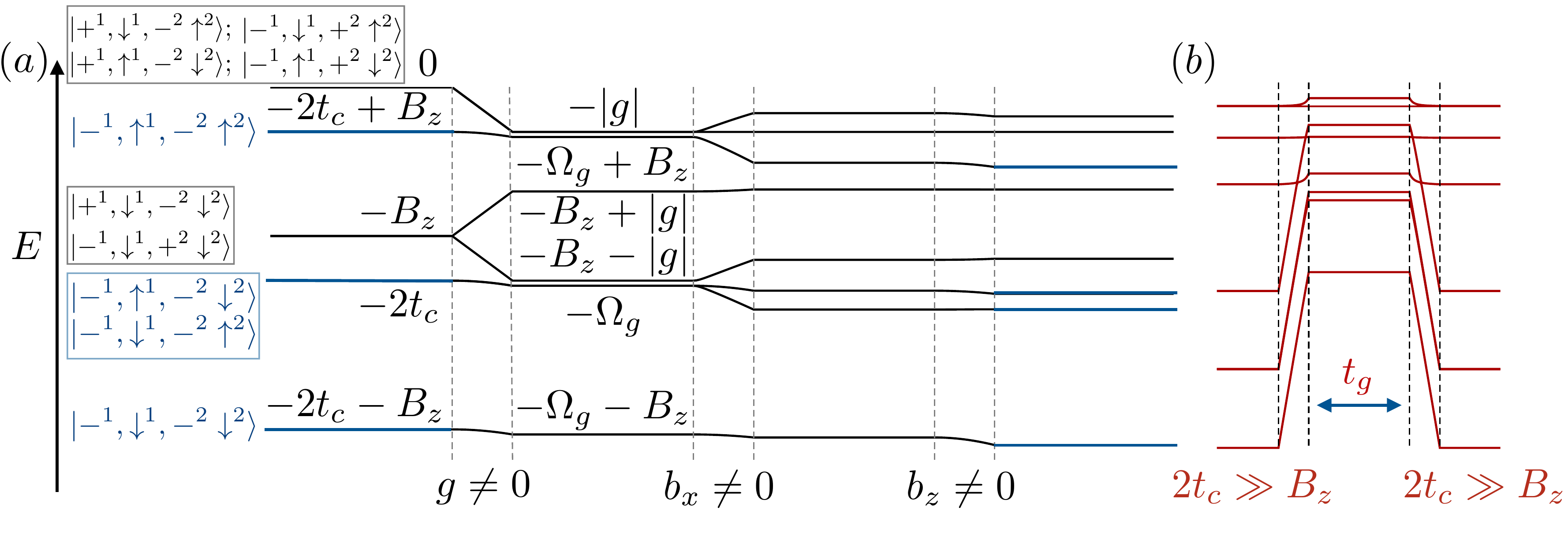}
	\caption{Lower energy levels corresponding to Hamiltonian \eqref{eq:H1}. (a) On the left, $g=\tilde{b}_x=\tilde{b}_z=0$. 
	The eigenstates are labeled by  $|\pm^1,\sigma^1,\pm^2,\sigma^2\rangle$, with $|\pm^i\rangle=(|R^i\rangle\pm|L^i\rangle)/\sqrt{2}$ and $\sigma\in \{\uparrow,\downarrow\}$.  The boxes in the leftmost part represent degenerate energy states and the four two-qubit states are indicated in blue color.
	Towards the right side of the figure, in the indicated regions, the different parameters $g$, $b_x$ and $b_z$ are turned on successively. (b) Low energy levels  during a possible realization of a two-qubit gate. First, the qubits do not interact because $2t_c\gg B_z$. Then, the tunnel coupling is reduced such that the qubits interact for a certain amount of time $t_g$, before $t_c$ is increased again, the qubits are uncoupled  and the desired quantum gate has been performed.}
	\label{Fig2}
\end{figure*}

Fig.~\ref{Fig2}(a) illustrates how the magnetic field gradients together with the dipole-dipole interaction generate a spin-spin interaction. On the left we show the lower energy levels of Hamiltonian \eqref{eq:H1} for the case $g=\tilde{b}_x=\tilde{b}_z=0$. 
The eigenstates can therefore be labeled as $|\pm^1,\sigma^1,\pm^2,\sigma^2\rangle$, where the first two labels correspond to DQD 1 and the last two to DQD 2, $|\pm^i\rangle=(|R^i\rangle\pm|L^i\rangle)/\sqrt{2}$, and $\sigma\in \{\uparrow,\downarrow \}$ is the spin component in the rotated axes.
In a first step we turn on the dipole-dipole interaction $g\neq0$ and observe an splitting between the levels at energy $-B_z$, possible due to the different charge distribution in both DQDs. Moreover, the orbital energy changes from $2t_c$ to $\Omega_g=\sqrt{4t_c^2+g^2}$ (see Appendix~\ref{sec:diagonalization}).
In the following step we turn on the transverse magnetic field gradient, 
which generates a splitting between  the two levels at energy $-\Omega_g$ due to the avoided crossing with the energy level at $-B_z-|g|$. Finally, the longitudinal gradient 
redistributes the energy levels and, as we show below, provides  a versatile set of two-qubit gates.

Within a perturbative approach in the magnetic field gradients, in particular under the conditions $b_{x}\ll(\Omega_g\pm g \pm B_z)$ and $b_{z}\ll\Omega_g \pm g$, we perform a Schrieffer-Wolff transformation \cite{Romhanyi2015,BRAVYI20112793} and decouple the states with  $|-^{1},\sigma^{1},-^{2},\sigma^{2}\rangle$ charge configuration 
from the rest (see Appendix~\ref{sec:SW}), obtaining an effective spin-spin Hamiltonian
\begin{equation}
\label{Heff}
H_{\rm eff}=\sum_{i=x,z}\left[J_{i0}\sigma_{i}^{(1)}+J_{0i}\sigma_{i}^{(2)}\right]
+\sum_{ij=x,z}J_{ij}\sigma_{i}^{(1)}\sigma_{j}^{(2)}\,,
\end{equation} 
where $J_{i0}$ and $J_{0i}$ correspond to the effective magnetic fields,
\begin{equation}
\label{eff-fields}
\begin{split}
J_{x0}&=-\frac{b_{x}b_{z}B_{z}[B_{z}^{2}-4(t_{c}^{2}+g^{2})]}{4[(B_{z}^{2}-4t_{c}^{2})^{2}-4B_{z}^2g^{2}]}\,,\\
J_{0z}=J_{z0}&=\frac{B_{z}}{2}+\frac{b_{x}^{2}B_{z}[B_{z}^{2}-4(t_{c}^{2}+g^{2})]}{4[(B_{z}^{2}-4t_{c}^{2})^{2}-4B_{z}^2g^{2}]}\,,
\end{split}
\end{equation}
and $J_{ij}$ to the spin-spin couplings, 
\begin{equation}
\label{couplings}
\begin{split}
J_{xx}&=\frac{4b_{x}^{2}t_{c}^{2}g}{[(B_{z}^{2}-4t_{c}^{2})^{2}-4B_{z}^2g^{2}]}\,,\\
J_{zx}&=\frac{b_{x}b_{z}g[B_{z}^{4}+32t_{c}^{4}- 4B_{z}^{2}(2t_{c}^{2}+g^{2})]}{8t_{c}^{2}[(B_{z}^{2}-4t_{c}^{2})^{2}-4B_{z}^2g^{2}]}\,,\\
J_{zz}&=\mp \frac{b_{z}^{2}g}{4t_{c}^{2}},
\end{split}
\end{equation}
where the  $\mp$ sign corresponds to the geometry Fig.~\ref{Fig1}(a) and ~\ref{Fig1}(b), respectively.  Moreover, for  geometry (a) we find  $J_{0x}=-J_{x0}$, $J_{xz}=-J_{zx}$, while in geometry (b)  $J_{0x}=J_{x0}$, $J_{xz}=J_{zx}$.  

At zero capacitive or Coulomb coupling, $g=0$, all the non-local couplings in Eq.~(\ref{couplings}) vanish, $J_{xx,xz,zz}=0$, leaving only finite  local terms.  The spin couplings  exhibit a strong dependence on the interdot tunnel coupling $t_{c}$ as it is evident in Eqs.~(\ref{couplings}) implying that they can be easily manipulated by means of voltage gates. This is illustrated in Fig.~\ref{Fig2}(b), where we show the lower energy levels for a non interacting situation ($2t_c\gg B_z$) on the left, the transition to a interacting situation by modification of $t_c$ for a time corresponding to some given gate time $t_g$  in the center, and back to a decoupled situation on the right.
The precise value of $t_c$ during the interaction phase and the gate time depends  on which quantum gate one wants to perform, as explained in the following section.

\section{Two-qubit gates}
\label{section2}
In this part we investigate the possible
  two-qubit gates that result from the time evolution under the physical Hamiltonian given by Eq.~(\ref{Heff}).\cite{PhysRevA.51.1015,PhysRevA.52.3457,loss2}
Two two-qubit gates are equivalent up to single-qubit operations when their Makhlin invariants coincide.\cite{Makhlin2002}
The two-qubit Hamiltonian (\ref{Heff}) is fairly general and the corresponding Makhlin invariants have been investigated numerically in the context of other type of QD spin qubits in  Refs.~\onlinecite{PhysRevB.91.035301} and  \onlinecite{PhysRevB.92.125409}.
Here, instead of using Eq.~(\ref{Heff}), we adopt a simplified version by making a rotating wave approximation (RWA), 
 \begin{equation}
\label{Heff2}
\bar{H}_{\rm eff}= J_{z0}\left(\sigma_{z}^{(1)}+\sigma_{z}^{(2)}\right)
+\frac{J_{xx}}{2}\sum_{i=x,y}\sigma_{i}^{(1)}\sigma_{i}^{(2)}
+J_{zz}\sigma_{z}^{(1)}\sigma_{z}^{(2)}\, .
\end{equation} 
The RWA is valid as long as $|J_{x0}|,|J_{xx}|,|J_{zx}|\ll J_{0z}$.

\begin{center}
\begin{table*}[t]
\begin{tabular}{ |c|c|c||c||c|c|}
 \hline
${\rm cos}(4J_{zz}t)$ &${\rm cos}(2J_{xx}t)$ & invariants& gate & gate time & coupling\\
 \hline
1  & -1    &$G_{1}=0$,\,$G_{2}=-1$& iSWAP& $t_g=\frac{\pi n}{2J_{zz}}=\frac{\pi (2m+1)}{2J_{xx}}$&$J_{zz}=\frac{n}{2m+1}J_{xx}$\\
-1  & -1    &$G_{1}=-1$,\,$G_{2}=-3$& SWAP& $t_g=\frac{\pi (2n+1)}{4J_{zz}}=\frac{\pi (2m+1)}{2J_{xx}}$&$J_{zz}=\frac{2n+1}{2(2m+1)}J_{xx}$\\
-1  & 1    &$G_{1}=0$,\,$G_{2}=1$& CNOT& $t_g=\frac{\pi (2n+1)}{4J_{zz}}=\frac{\pi m}{J_{xx}}$&$J_{zz}=\frac{1}{4}\frac{2n+1}{m}J_{xx}$\\
0  & 0    &$G_{1}=\pm i/4$,\,$G_{2}=0$& $\sqrt{\text{SWAP}}$& $t_g=\frac{\pi (2n+1)}{8J_{zz}}=\frac{\pi (2m+1)}{4J_{xx}}$&$J_{zz}=\frac{1}{2}\frac{2n+1}{2m+1}J_{xx}$\\
1  & 0    &$G_{1}=1/4$,\,$G_{2}=1$& $\sqrt{\text{iSWAP}}$& $t_g=\frac{\pi n}{2J_{zz}}=\frac{\pi (2m+1)}{4J_{xx}}$&$J_{zz}=2\frac{n}{2m+1}J_{xx}$\\
 \hline
\end{tabular}
\caption{Values of Makhlin's invariants obtained from Eqs.~(\ref{Invariants2}). The fifth column (gate time) contains the times at which the values of the invariants for the respective two-qubit gates (fourth column) are achieved,  where $n$ and $m$ are integers. The last column indicates which relation the couplings must fulfill to obtain the respective two-qubit gate.}
\label{Table}
\end{table*}
\end{center}
In the following, we evaluate Makhlin's invariants of the time-evolution generated by the Hamiltonian given by Eq.~(\ref{Heff2}) and compare them with the well-known two-qubit gates. The invariants can be obtained from the following expressions\cite{Makhlin2002}
\begin{equation}
\label{makhlin}
\begin{split}
G_{1}(t)&=\frac{{\rm Tr}^{2}m(t) {\rm det}U^{\dagger}(t)}{16 }\,,\\
G_{2}(t)&=\frac{[{\rm Tr}^{2}m(t)-({\rm Tr}\,m(t))^{2}]{\rm det} U^{\dagger}(t)}{4}\,,\\
\end{split}
\end{equation}
where $U(t)={\rm e}^{-i\bar{H}_{\rm eff}t}$ is the unitary time evolution operator of the effective Hamiltonian $\bar{H}_{\rm eff}$,  $m(t) = M^{T}_{\rm B}(t)M_{\rm B}(t)$, and $M_{\rm B}(t)=Q^{\dagger}U(t)Q$ describes the  transformation of $U(t)$ into the Bell basis, where
\begin{equation}
\label{Bell}
Q=
\begin{pmatrix}
1&0&0&i\\
0&i&1&0\\
0&i&-1&0\\
1&0&0&-1
\end{pmatrix}\,.
\end{equation}

\begin{figure}[!b]
	\centering
	\includegraphics[width=\columnwidth]{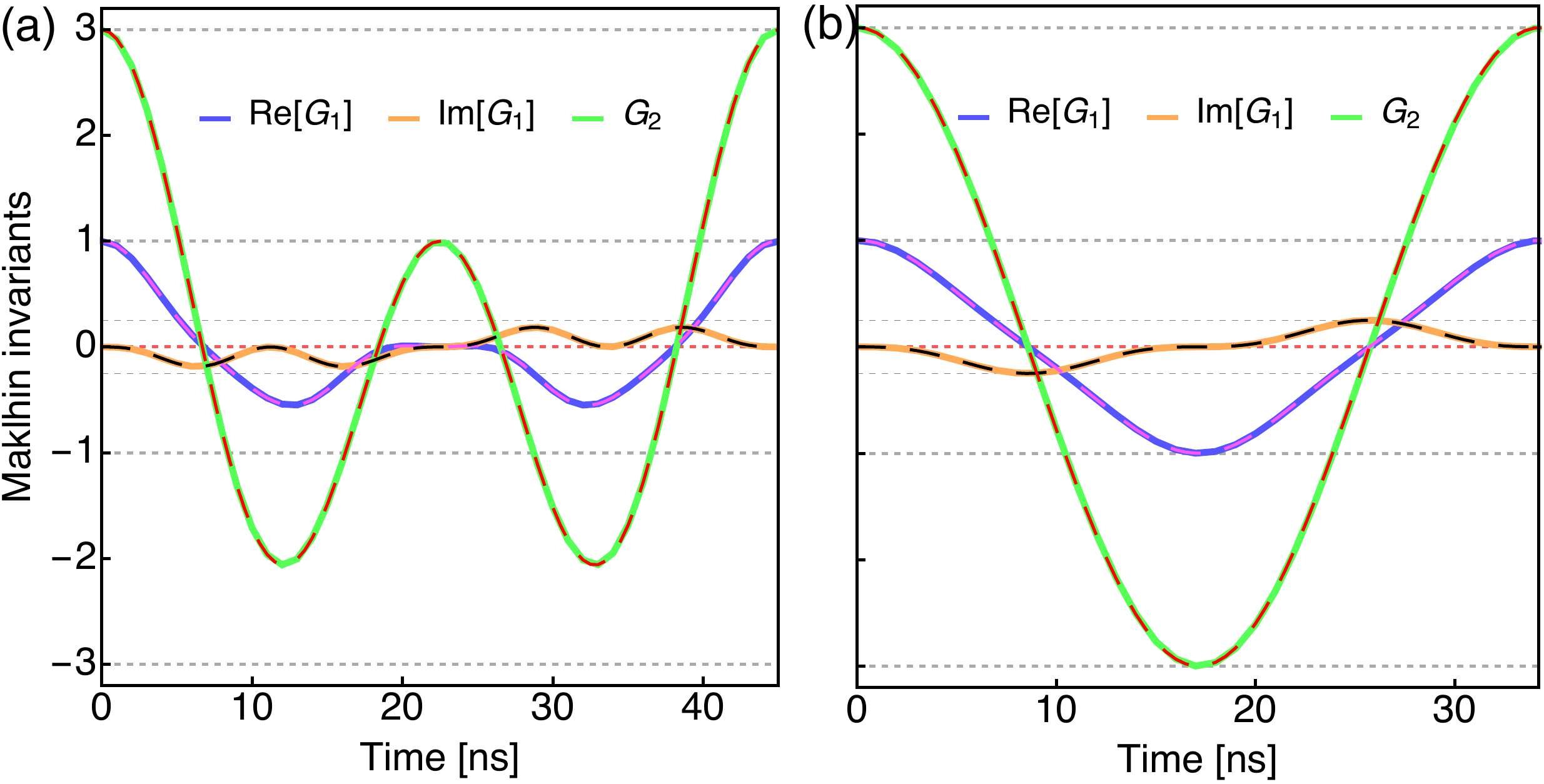}
	\caption{Time evolution of the Makhlin invariants depicting the realisation of multiple two-qubit gates.  Thick solid curves correspond to the exact invariants calculated for the Hamiltonian given by Eq.\,(\ref{Heff}), while dashed curves inside them represent the invariants calculated by using the analytical expressions given by Eqs.\,(\ref{Invariants2}) within RWA. 
	(a) The invariants for a CNOT gate correspond to $G_{1}=0$, $G_{2}=1$, reached in the center of the plot.
	 (b) The invariants for a SWAP gate correspond to $G_{1}=-1$, $G_{2}=-3$, reached in the center of the plot. The $\sqrt{\rm SWAP}$ gate occurs at this half time.
	 Parameters:  $B_z=24\,\mu\text{eV}$, $g=-25\,\mu\text{eV}$, $b_x=2\,\mu\text{eV}$, and $b_z=1\,\mu\text{eV}$, and (a) $t_{c}=16.5\mu\text{eV}$ (b)  $t_{c}=14.4\mu\text{eV}$. 
	}
	\label{Fig1a}
\end{figure}

Within the RWA, we obtain simple analytical expressions for the invariants,
\begin{equation}
\label{Invariants2}
\begin{split}
G_{1}(t)&=\frac{{\rm e}^{-4iJ_{zz}t}}{4}\Big\{ 1+{\rm e}^{4iJ_{zz}t} 
  {\rm cos}(2J_{xx}t)\Big\}^{2}\,,\\
G_{2}(t)&=  2 {\rm cos}(2J_{xx}t)+{\rm cos}(4J_{zz}t)\,,
\end{split}
\end{equation}
and it is easy to analyze them and derive conditions to reach  the values that correspond to the well-known two-qubit gates. After some algebra, we find that the physical Hamiltonian, given by Eq.~(\ref{Heff2}), allows the realization of multiple two-qubit gates under a set of conditions for the relation between $J_{xx}$ and $J_{zz}$, as  described in detail in Table \ref{Table}. Given that those couplings can be externally modified at will, this demonstrates that the capacitively coupled flopping-mode spin qubits represent a very versatile platform for realizing two-qubit gates.

In order to visualize the values of the invariants obtained in Table \ref{Table}, we  plot the  time-dependence of both the exact  and the approximated result given by  Eqs.~(\ref{Invariants2}) for different values of the tunnel coupling $t_c$ in Fig.~\ref{Fig1a} which results in different qubit-qubit couplings. 
 For instance, in panel (a)  $J_{xx}=4 J_{zz}$ and $J_{zz}=0.5$ such that a two-qubit gate equivalent to CNOT is realized at $t_g=\pi/(J_{xx})$ where the invariants are $G_{1}=0$ and $G_{2}=1$. In (b)   $J_{xx}=4 J_{zz}$ and we have two situations: first, at $t_g=\pi/(4 J_{xx})$ the invariants read $G_{1}=-i/4$ and $G_{2}=0$, giving rise to $\sqrt{\text{SWAP}}$, and this periodically repeats; second at   $t_g=\pi/(2J_{xx})$ $G_{1}=-1$ and $G_{2}=-3$ which corresponds to SWAP.

\section{Dephasing and fidelity}
\label{Fidelity}
In the previous section we have demonstrated that multiple two-qubit gates can be realized with the physical Hamiltonian for  capacitively coupled flopping-mode spin qubits. 
However, this spin qubit inherently  suffers from decoherence even in magnetic-noise free materials, since it possesses  an  electric dipole that exposes it to charge noise.
To model charge-noise induced decoherence we introduce two independent bosonic baths for the two DQDs, with bosonic annihilation operators $a_{ik}$  for DQD $i=1,2$.
Given the electron-bath coupling rates $\lambda_{ik}$, the interaction between the electrons and these baths can be written as $V=\sum_{i=1,2}\sum_k \lambda_{ik} \tau_z^{(i)} (a_{ik}+a_{ik}^{\dagger})$.
Applying the same basis transformations explained in Appendix~\ref{sec:diagonalization} to this Hamiltonian and the SW transformation presented in Appendix~\ref{sec:SW}, we find that, in terms of the perturbed (or dressed) qubit operators, this electron-bath coupling Hamiltonian  reads
\begin{equation}
V_{\rm eff} = \sum_{i,j,k}\sum_{q=x,z} \lambda_{ik}  
\mu_{ijq}\sigma_{q}^{(j)} (a_{ik}^{\dagger}+a_{ik}), \label{eq:Veff}
\end{equation}
where 
$\mu_{ijq}=\alpha_{q}^{(i)}\delta_{ij}+\beta_{q}^{(i)}\delta_{i\bar{j}}$
and
$\delta_{i\bar{j}}=1-\delta_{ij}$.
The transverse magnetic field gradient couples the qubits transversally to the bath via
\begin{equation}
\label{alphabeta}
\begin{split}
\alpha_{x}^{(1)}&=\alpha_{x}^{(2)} = \frac{4b_x t_c^2\left[B_z^2-2(2t_c^2+g^2)\right]}{\sqrt{4t_c^2+g^2}\left[(B_z^2-4t_c^2)^2-4B_z^2g^2\right]}\,,\\
\beta_{x}^{(1)}&=\beta_{x}^{(2)}=\frac{8b_x t_c^2g}{(B_z^2-4t_c^2)^2-4B_z^2g^2}\, ,
\end{split}
\end{equation}
contributing to relaxation, while the longitudinal gradient couples them longitudinally, giving rise to pure dephasing
\begin{equation}
\label{alphabeta}
\begin{split}
\alpha_{z}^{(1)}&=\mp\alpha_{z}^{(2)}=-\frac{b_z(2t_c^2+g^2)}{2t_c^2\sqrt{4t_c^2+g^2}}\,,\\
\beta_{z}^{(1)}&=\mp\beta_{z}^{(2)}=\mp\frac{b_zg}{2t_c^2}\, .
\end{split}
\end{equation}
Here the upper and lower signs correspond to the horizontal and vertical  geometries shown in Fig.~\ref{Fig1} (a) and (b), respectively.

\begin{figure}[!t]
	\centering
	\includegraphics[width=.4\textwidth]{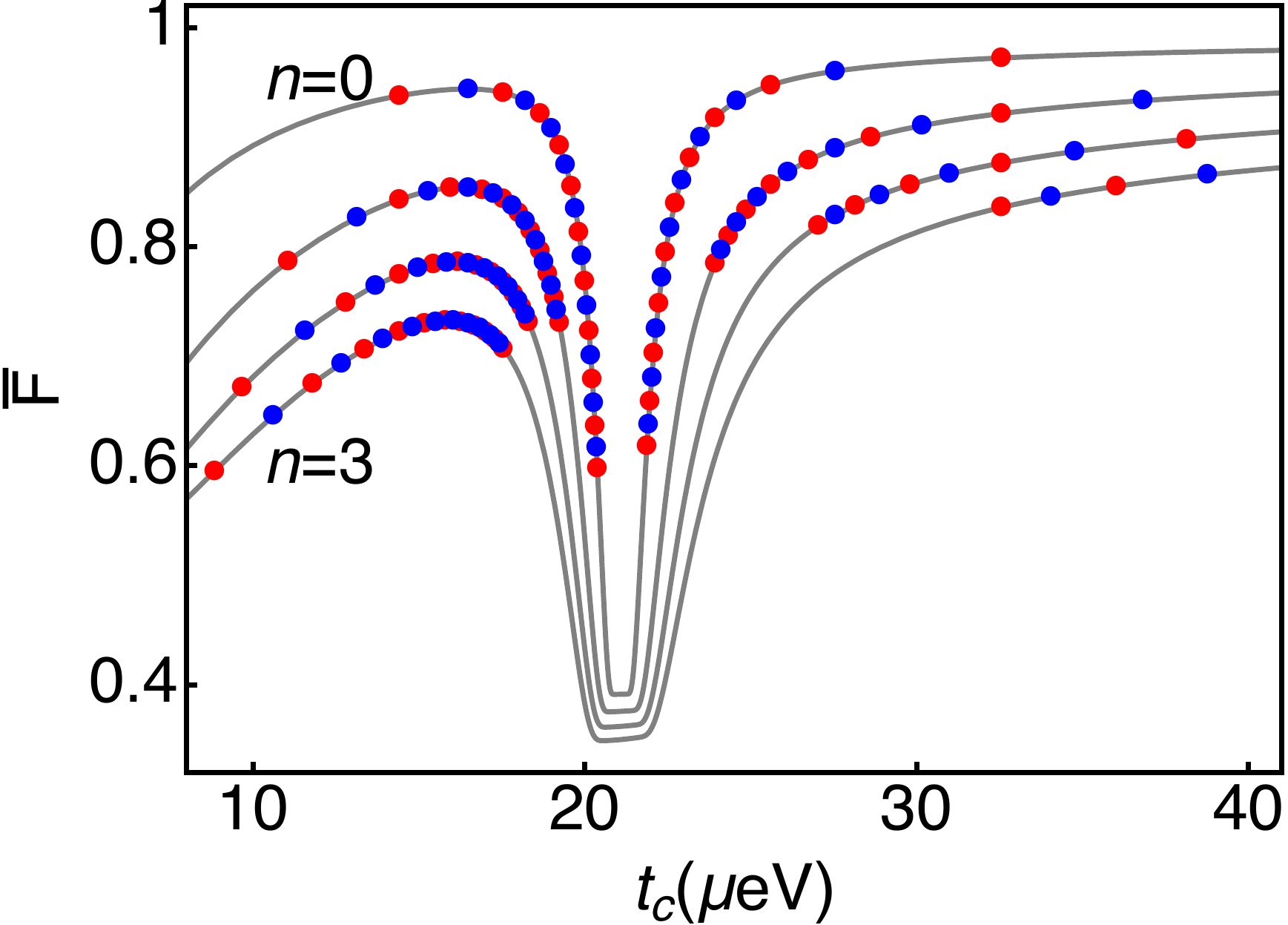}
	\caption{Average fidelity $\bar{F}$ of a SWAP (red dots) and a CNOT (blue dots)  gate between two spin qubits as a function of the tunnel coupling $t_c$. Gray curves represent the fidelity  between the result of the evolution given by Eq.\eqref{eq:mastereq} and the one  given by $\bar{H}_{\rm eff}$ during the time $t_g=(2n+1)\pi/(4J_{zz})$ and plotted only for reference.	The rest of parameters are $\gamma_{-,1}=\gamma_{-,2}=0.02\,\mu\text{eV}$, $2\gamma_{z,1}=2\gamma_{z,2}=0.4\,\mu\text{eV}$, $B_z=24\,\mu\text{eV}$, $g=-25\,\mu\text{eV}$, $b_x=2\,\mu\text{eV}$, and $b_z=1\,\mu\text{eV}$. The SWAP gate is performed when $(2n+1)J_{xx}=2(2m+1)J_{zz}$, while  the CNOT gate is performed when $(2n+1)J_{xx}=4mJ_{zz}$. We have considered only the shortest gate times $t_g=(2n+1)\pi/(4J_{zz})$ for $n = 0, 1, 2, 3$.}
	\label{Fig4}
\end{figure}

Tracing over the  bath degrees of freedom, within the Born-Markov approximation and assuming zero temperature, we find the following master equation for the partial density matrix for the two-qubit system
\begin{equation}
\dot{\rho}=-i \left[\bar{H}_{\rm eff},\rho\right]
+\sum_{i,q=-,z} \gamma_{q,i}\mathcal{D}[\sum_{j}\mu_{ijq}\sigma_{q}^{(j)}]\rho\ ,
\label{eq:mastereq}
\end{equation}
with the dissipation superoperator $\mathcal{D}[A]\rho = A\rho A^\dagger-\frac{1}{2}A^\dagger A\rho-\frac{1}{2}\rho A^\dagger A$, $\alpha_-^{(i)}=\alpha_x^{(i)}$ and $\beta_-^{(i)}=\beta_x^{(i)}$.
Here, the rates $\gamma_{-,i}$ and $2\gamma_{z,i}$ correspond to the charge relaxation and pure dephasing rates in DQD $i=1,2$. Based on recent experiments in DQDs,~\cite{Mi2018}  the relaxation rates have been chosen to be $\gamma_{-i}= 
0.02\,\mu\text{eV}\sim 2\pi\times 4.8\,\text{MHz}$ and the pure dephasing rates $2 \gamma_{zi}= 0.4\,\mu\,\text{eV}\sim 2\pi\times 96.7\,\text{MHz}$.
%
%
From the simple master given by Eq.\,(\ref{eq:mastereq}) we can then compute the average two-qubit gate fidelity $\bar{F}$ as in Refs.\,[\onlinecite{PhysRevA.71.062310}] and [\onlinecite{PhysRevB.96.115407}] by comparing the resulting mixed state and the targeted pure state and averaging over all possible pure initial states.

In Fig.~\ref{Fig4}, we present $\bar{F}$ as a function of the interdot tunnel coupling $t_{c}$. The gray lines represent  $\bar{F}$ between the result of the evolution given by Eq.\eqref{eq:mastereq} and the one given by $\bar{H}_{\rm eff}$ during the time $t_g=(2n+1)\pi/(4J_{zz})$ and serve as a guide to the eye.
The red and blue dots show the SWAP and CNOT average gate fidelities for different values of $t_{c}$ which correspond to different integers $\{n,m\}$ satisfying the condition  $J_{zz}=(2n+1)J_{xx}/[2(2m+1)]$  and $J_{zz}=(2n+1)J_{xx}/(4m)$, respectively. Note that the red and blue points correspond to selected points on the gray lines where the evolution is actually an interesting two-qubit gate.  The drop in $\bar{F}$ at around $t_c\approx 21\,\mu\text{eV}$ corresponds to the regime where $\Omega_g\approx B_z+|g|$ (see Fig. 2), in which the spin qubit suffers severely from charge noise.
Given the high-fidelity single-qubit gates reported for quantum dot spin qubits,~\cite{Yoneda2018} we have not accounted here for imperfections in the single-qubit gates necessary to map the physical evolution to the targeted two-qubit gate.
Analogously to these exemplary two-qubit gates, the other quantum gates can also be realized with a fidelity close to 95$\%$.  These values can easily increase with a more optimistic choice of the pure dephasing rates, as observed in other similar experiments.\cite{Woerkom2018}

\section{Alternative geometry with two micromagnets}
\label{alt}
\begin{figure}[h]
	\centering
	\includegraphics[width=\columnwidth]{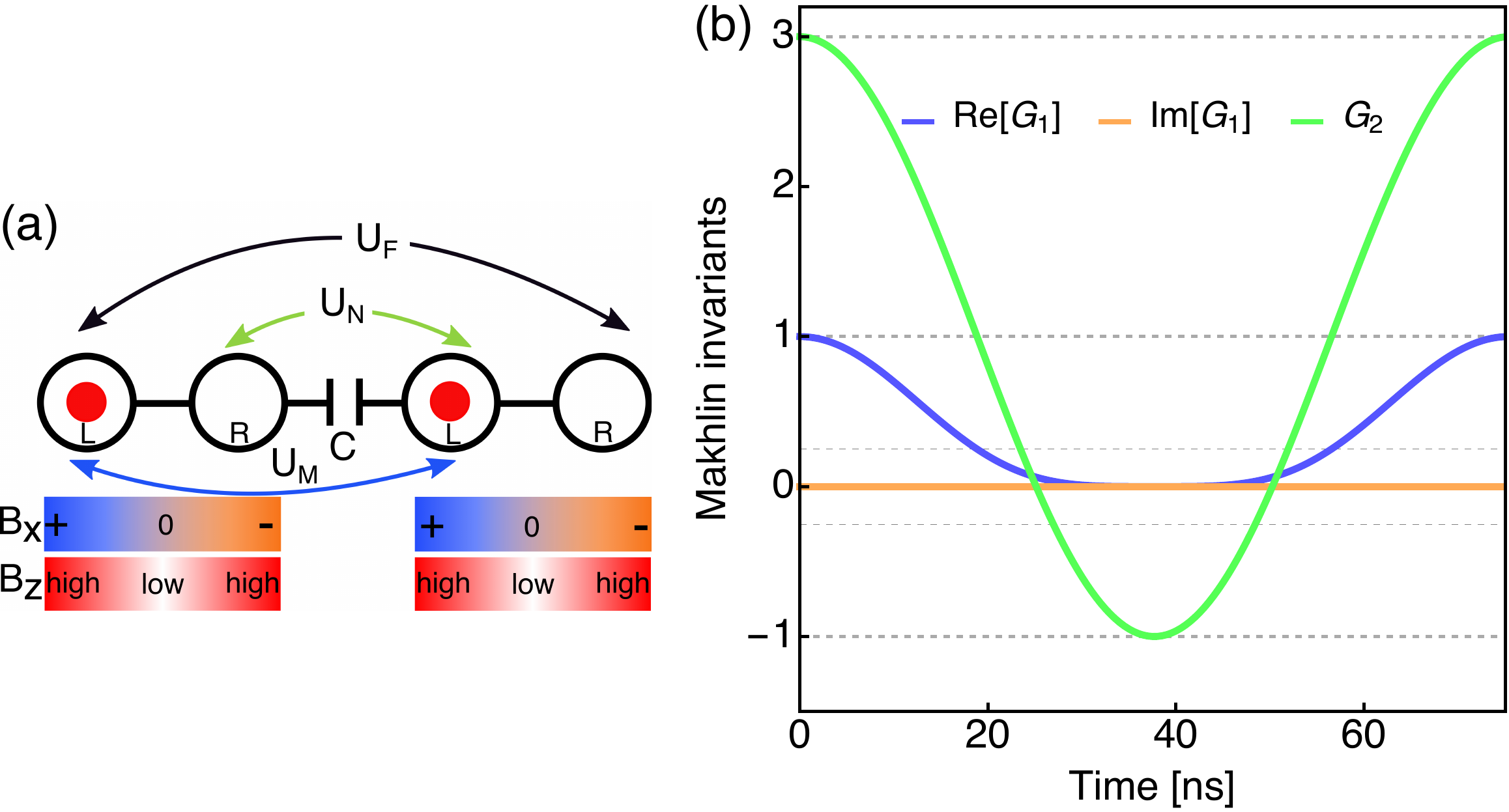}
	\caption{(a) Alternative geometry where each DQD is centred in the micromagnet stray field. (b) Time evolution of the Makhlin invariants depicting the realization of  iSWAP and $\sqrt{\text{iSWAP}}$ two-qubit gates, obtained from Eqs.~(\ref{maklhinGeo2}) for the alternative geometry with two micromagnets at zero detuning. The invariants for an iSWAP gate correspond to $G_{1}=0$, $G_{2}=-1$, reached in the center of the plot. The invariants for a $\sqrt{\text{iSWAP}}$ gate correspond to $G_{1}=1/4$, $G_{2}=1$, reached at half this time.
	 Parameters:  $B_z=24\,\mu\text{eV}$, $g=-25\,\mu\text{eV}$, $b_x=2\,\mu\text{eV}$, and $t_{c}=16.5\mu\text{eV}$.  
	}
	\label{Fig5}
\end{figure}

Another possible array of flopping-mode spin qubits could be fabricated by centering  each DQD on the micromagnet stray field, in such a way that both $\tilde{B}_x$ and $\tilde{b}_z$ in Eq. \eqref{Zeeman1} are zero, as shown in Fig.\,\ref{Fig5}(a).
In this situation, the couplings and effective magnetic fields can be directly obtained by substituting $b_{z}=0$, $B_{z}=\tilde{B}_{z}$ and $b_x=\tilde{b}_{x}$ in Eqs.\,(\ref{eff-fields}) and (\ref{couplings}) which leads to an effective Hamiltonian where the only finite terms are $J_{0z}$ and $J_{xx}$. In this case, we can calculate the exact Makhlin invariants which, after some algebra, read
\begin{equation}
\label{maklhinGeo2}
\begin{split}
G_{1}(t)&=\frac{[4J_{0z}^{2}+Q^{2}{\rm cos}(2J_{xx}t)+J_{xx}^{2}{\rm cos}(2Qt)]^{2}}{4Q^{4}}\,,\\
G_{2}(t)&=1+\frac{2{\rm cos}(2J_{xx}t)[4J_{0z}^{2}+J_{xx}^{2}{\rm cos}(2Qt)]}{Q^{2}}\,,\\
\end{split}
\end{equation}
where $Q=\sqrt{4J_{0z}^{2}+J_{xx}^{2}}$.

The zero  $J_{zz}$ coupling obtained here has profound consequences. A simple inspection of Eqs.~(\ref{maklhinGeo2})  suggests that there are two important situations for the Makhlin invariants. First, for ${\rm cos}(2Qt)=1$ and ${\rm cos}(2J_{xx}t)=0$ one finds $G_{1}=1/4$ and $G_{2}=-1$, which corresponds to a $\sqrt{\text{iSWAP}}$ two-qubit gate. Second,  ${\rm cos}(2Qt)=1$ and ${\rm cos}(2J_{xx}t)=-1$ yields $G_{1}=0$ and $G_{2}=1$, which corresponds to a iSWAP two-qubit gate.  The conditions for the expected times and ratios between the couplings can be obtained as carried out in the previous section. We then obtain for $\sqrt{\text{iSWAP}}$,
\begin{equation}
t_g=\frac{\pi (2m+1)}{4J_{xx}}\,,\quad J_{xx}=\frac{2m+1}{n}\frac{Q}{4}\,,
\end{equation}
while for iSWAP,
\begin{equation}
t_g=\frac{\pi (2m+1)}{2J_{xx}}\,,\quad J_{xx}=\frac{2m+1}{n}\frac{Q}{2}\,,
\end{equation}
In order to clarify the discussion above, in Fig.~\ref{Fig5}(b) we plot the Makhlin invariants from Eqs.~(\ref{maklhinGeo2})  as a function of time, where we can clearly observe that they acquire the values for $\sqrt{\text{iSWAP}}$ and iSWAP at  $t_g\approx 18.8$\,ns and $t_g\approx 37.7$\,ns, respectively, for the chosen realistic parameters.
Note that operating the qubits at a different detuning, $\epsilon_2=-\epsilon_1\neq (U_N-U_F)/2$, results in a finite $J_{zz}$ and other two-qubit gates are also realizable in a single step.

To close this part, we point out that since in this case $b_z=0$, the noise model used in Sec.\,\ref{Fidelity} would imply that there is no pure dephasing, an effect that overestimates the fidelity. Thus, a noise model accounting for coupling of charge noise to higher order in the fluctuating parameters  and other sources of error, such as as leakage to other energy levels during the operation, is required.  This is, however, beyond of the scope of this work.

\section{Conclusions}
\label{conclusions}
We have investigated the realization of single-step two-qubit gates in capacitively coupled flopping-mode spin qubits. In particular, by calculating the Makhlin invariants from the  physical Hamiltonian for the qubit-qubit interaction, we have demonstrated that multiple two-qubit quantum gates can be realized with the same setup by using realistic experimental parameters.~\cite{Neyens2019,Mi2018,Samkharadze2018} Interestingly, a variety of two-qubit gates can be realized with fidelity higher than $95\%$, even accounting for dephasing.

In the first part, we have considered that the two flopping-mode spin qubits are embedded in a single micromagnet stray field, while in the second part each DQD is centered in a micromagnet stray field.
In both schemes, the two-qubit gates  emerge as a result of the capacitive coupling, via Coulomb interaction. 
Due to an interplay between the transverse $J_{xx}$ and the longitudinal $J_{zz}$ coupling, the former setup exhibits the realization of multiple gates, e.g. the CNOT \cite{PhysRevA.52.3457} and $\sqrt{\text{SWAP}}$ \cite{loss2}, which together with single qubit rotations allows for  universal quantum computation.\cite{PhysRevA.51.1015} The latter geometry, on the other hand,  permits the realization of iSWAP and $\sqrt{\text{iSWAP}}$ in a single step, attributed to the vanishing  $J_{zz}$ coupling. These gates combined with single-qubit gates are also universal.\cite{Imamoglu1999,Schuch2003}

Interestingly,  long-distance coupling between these types of qubits, mediated by superconducting resonator photons, was recently experimentally  realized.~\cite{Borjans2019}
Our work, therefore, demonstrates that capacitively coupled flopping-mode spin qubits constitute a promising platform for scalable quantum computation nodes.

\begin{acknowledgments}
We thank A. Black-Schaffer, V. Shkolnykov, and E. Sj\"{o}qvist for helpful discussions.
This work was financially supported by ARO grant through grant no.~W911NF-15-1-0149
and DFG through the collaborative research center SFB 767.  
J. C. acknowledges support from C.F. Liljewalchs stipendiestiftelse Foundation.
\end{acknowledgments}

\appendix

\section{Basis transformations} 
\label{sec:diagonalization}
In this appendix we show the basis transformations used to account for the dipole-dipole interaction exactly and obtain the transformed Hamiltonian that serves as starting point for the perturbative treatment of the magnetic field gradients.
First of all, we rotate the spin quantization axis in each QD by applying the transformation 
\begin{equation*}
U_{\sigma}=\left(
\begin{array}{cc}
 \cos \frac{\theta}{2} & -\sin \frac{\theta}{2} \\
 \sin \frac{\theta}{2} & \cos \frac{\theta}{2} \\
\end{array}
\right)\otimes \left(
\begin{array}{cc}
 \cos \frac{\theta}{2} & \mp\sin \frac{\theta}{2} \\
\pm \sin \frac{\theta}{2} & \cos \frac{\theta}{2} \\
\end{array}
\right)\otimes \mathbb{1}\otimes \mathbb{1} \, ,
\end{equation*}
where $\theta=\arctan{(\tilde{B}_x/\tilde{B}_z)}$. Here and in the following the subspaces order is $\sigma^{(1)}\otimes \sigma^{(2)}\otimes \tau^{(1)}\otimes \tau^{(2)}$.
The transformed Zeeman Hamiltonian, $ H'_{\rm z}=U_{\sigma}^{\dagger}H_{\rm z} U_{\sigma}$, reads
\begin{equation*}
\begin{split}
    H'_{\rm z}&=\frac{1}{2}\sum_{i=1,2}\Big[B_{z}\sigma_{z}^{(i)}+(\mp 1)^{i-1}b_{z}\sigma_{z}^{(i)}\tau_{z}^{(i)}
+
b_{x}\sigma_{x}^{(i)}\tau_{z}^{(i)}\Big] \, ,
\end{split}
\end{equation*}
while the rest of the terms of the Hamiltonian are unchanged.

Given the detuning choice explained in the main text, it is natural to perform another basis transformation, 
\begin{equation}
U_{\tau}=\frac{1}{2}\mathbb{1}\otimes \mathbb{1}\otimes \left(
\begin{array}{cc}
 1 & -1 \\
 1 &1 \\
\end{array}
\right)\otimes \left(
\begin{array}{cc}
 1 & -1 \\
 1 &1 \\
\end{array}
\right)  \, ,
\end{equation}
which, together with $U_{\sigma}$, transforms   Eq.\,\eqref{eq:H1} into
\begin{equation*}
\begin{split}
    H'&=(U_{\sigma}U_{\tau})^{\dagger}H (U_{\sigma}U_{\tau})=\sum_{i=1,2}t_c\tau_{z}^{(i)}+g\tau_x^{(1)}\tau_x^{(2)}
    \\&+\frac{1}{2}\sum_{i=1,2}\Big[B_{z}\sigma_{z}^{(i)}-(\mp 1)^{i-1}b_{z}\sigma_{z}^{(i)}\tau_{x}^{(i)}
-
b_{x}\sigma_{x}^{(i)}\tau_{x}^{(i)}\Big] \, .
\end{split}
\end{equation*}

Finally, we apply a transformation that exactly diagonalizes the dipole-dipole coupling term,
\begin{equation*}
V_{\tau}=\mathbb{1}\otimes \mathbb{1}\otimes \left(
\begin{array}{cccc}
 \cos{(\phi/2)}  & 0 & 0 & -\sin{(\phi/2)} \\
  0 & 1/\sqrt{2} & -1/\sqrt{2} & 0 \\
   0 & 1/\sqrt{2} & 1/\sqrt{2} & 0 \\
    \sin{(\phi/2)}  & 0 & 0 & \cos{(\phi/2)}  
\end{array}
\right)  \, ,
\end{equation*}
where $\phi=\arctan{(g/2t_c)}$. With this, the final Hamiltonian $H''=V_{\tau}^{\dagger}H' V_{\tau}$ reads
\begin{equation}
\begin{split}
    H''&=\frac{\Omega_g+g}{2}\tau_{z}^{(1)}+\frac{\Omega_g-g}{2}\tau_{z}^{(2)}\\
&   +\frac{1}{2}\sum_{i=1,2}B_{z}\sigma_{z}^{(i)}+W\, . \label{eq:Hpp}
\end{split}
\end{equation}
where $\Omega_g=\sqrt{4t_c^2+g^2}$ and $W$ contains all the terms induced by the magnetic field gradients and will be treated in the following as a perturbation,
\begin{equation}
\begin{split}
    W=&-\frac{b_x}{2\sqrt{2}}
\left(\cos{\frac{\phi}{2}}-\sin{\frac{\phi}{2}}\right)\left(\sigma_{x}^{(1)}\tau_{x}^{(1)}-\sigma_{x}^{(2)}\tau_{x}^{(1)}\tau_{z}^{(2)}\right)\\
&-\frac{b_x}{2\sqrt{2}}
\left(\cos{\frac{\phi}{2}}+\sin{\frac{\phi}{2}}\right)\left(\sigma_{x}^{(2)}\tau_{x}^{(2)}+\sigma_{x}^{(1)}\tau_{z}^{(1)}\tau_{x}^{(2)}\right)\\
&-\frac{b_z}{2\sqrt{2}}
\left(\cos{\frac{\phi}{2}}-\sin{\frac{\phi}{2}}\right)\left(\sigma_{z}^{(1)}\tau_{x}^{(1)}\mp\sigma_{z}^{(2)}\tau_{x}^{(1)}\tau_{z}^{(2)}\right)\\
&-\frac{b_z}{2\sqrt{2}}
\left(\cos{\frac{\phi}{2}}+\sin{\frac{\phi}{2}}\right)\left(\pm\sigma_{z}^{(2)}\tau_{x}^{(2)}+\sigma_{z}^{(1)}\tau_{z}^{(1)}\tau_{x}^{(2)}\right)\, . \label{eq:W}
\end{split}
\end{equation}
This perturbation indicates under which conditions the treatment is valid, since the off-diagonal terms need to be smaller as compared to the energy differences, i.e., $b_x\ll \Omega_g\pm g\pm B_z$ and $b_z\ll \Omega_g\pm g$. 

The electron-bath Hamiltonian $V$ given in the main text can also be transformed to this new basis,
\begin{equation}
\begin{split}
V''&=(U_{\sigma}U_{\tau}V_{\tau})^{\dagger}V (U_{\sigma}U_{\tau}V_{\tau})
\\&=-\sum_{k}\frac{\lambda_{1,k}}{\sqrt{2}}\left[\tau_x^{(1)}\left(\cos{\frac{\phi}{2}}-\sin{\frac{\phi}{2}}\right)\right. 
\\&\left. +\tau_z^{(1)}\tau_x^{(2)}\left(\cos{\frac{\phi}{2}}+\sin{\frac{\phi}{2}}\right)\right](a_{1,k}^{\dagger}+a_{1,k})
\\&-\sum_{k}\frac{\lambda_{2,k}}{\sqrt{2}}\left[\tau_x^{(2)}\left(\cos{\frac{\phi}{2}}+\sin{\frac{\phi}{2}}\right)\right. 
\\&\left. -\tau_x^{(1)}\tau_z^{(2)}\left(\cos{\frac{\phi}{2}}-\sin{\frac{\phi}{2}}\right)\right](a_{2,k}^{\dagger}+a_{2,k}) \, .
\end{split}
\end{equation}
As apparent from this expression, the bath surrounding one of the DQDs  can induce charge relaxation via both $\tau_-^{(1)}$ and $\tau_-^{(2)}$.

\section{Schrieffer-Wolff transformation}
\label{sec:SW}
The Schrieffer-Wolff (SW) transformation allows to decouple two weakly coupled subspaces to a desired order in the weak perturbative parameter. This transformation also allows to treat different perturbative parameters, as is the case in our problem, with the Hamiltonian given in Eqs.~\eqref{eq:Hpp} and \eqref{eq:W}. Given the two subspaces or blocks A and B, one separates first the interaction $W$ in a term that couples them, block-off-diagonal term $W_2$, and a block-diagonal term, $W_1$. The rest of the terms are diagonal, $H_{\rm d}$. Applying the transformation matrix $S$, with $\left[H_d,S\right]=-W_2$, to the Hamiltonian, one obtains the second-order Hamiltonian 
\begin{equation}
H_{\rm SW}=H_{\rm d}+W_1+\frac{1}{2}\left[W_2,S\right]
\end{equation}
which is block-diagonal. A projection into the interesting subspace gives therefore an effective Hamiltonian of reduced dimension. Although this procedure can be performed to the desired order~\cite{Romhanyi2015}, we restrict ourselves here to  second order.    In our case, the goal is to find an effective Hamiltonian for the subspace with $\tau_z^{(i)}=-1$, with diagonal terms $\{-\Omega_g-B_z,-\Omega_g,-\Omega_g,-\Omega_g+B_z\}$. The result is the two-qubit Hamiltonian in Eq.~\eqref{Heff}, which is now written in the basis dressed by the SW transformation.

The electron-bath Hamiltonian, $V''$, also has to be expressed in the dressed basis, which reads
\begin{equation}
V_{\rm SW}=V''+\left[V'',S\right] \ .
\end{equation}
Then, the result can be projected into the desired subspace obtaining $V_{\rm eff}$ in Eq.~\eqref{eq:Veff}.

\bibliography{biblio}

\end{document}